\documentclass[twocolumn, pra, showpacs,superscriptaddress]{revtex4}
\usepackage{amsfonts}
\usepackage{amssymb}
\usepackage{graphicx}
\usepackage{dcolumn}
\usepackage{bm}
\usepackage{amsmath}

\begin{document}

\title{Tavis-Cummings model beyond the rotating wave approximation:
Inhomogeneous coupling}
\author{Lijun Mao}
\affiliation{Institute of Theoretical Physics, Shanxi University, Taiyuan 030006, P. R.
China}
\author{Sainan Huai}
\affiliation{Institute of Theoretical Physics, Shanxi University, Taiyuan 030006, P. R.
China}
\author{Yunbo Zhang}
\email{ybzhang@sxu.edu.cn}
\affiliation{Institute of Theoretical Physics, Shanxi University, Taiyuan 030006, P. R.
China}

\begin{abstract}
We present the analytical solution of the Tavis-Cummings (TC) model for more
than one qubit inhomogeneously coupled to a single mode radiation field
beyond the rotating-wave approximation (RWA). The significant advantage of
the displaced oscillator basis enables us to apply the same truncation
techniques adopted in the single qubit Jaynes-Cummings (JC) model to the
multiple qubits system. The derived analytical spectrum match perfectly the
exact diagonalization numerical solutions of the inhomogeneous TC model
in the parameter regime where the qubits transition frequencies are far
off-resonance with the field frequency and the interaction strengths reach
the ultra-strong coupling regime. The two-qubit TC model is quasi-exactly
solvable because part of the spectra can be determined exactly in the
homogeneous coupling case with two identical qubits or with symmetric(asymmetric) 
detuning. By means of the fidelity of quantum states we identify several
nontrivial level crossing points in the same parity subspace, which implies that 
homogeneous coupled two-qubit TC model with $\omega_1=\omega_2$ or 
$\omega_1\pm\omega_2=2\omega_c$ is integrable.
We further explore the time evolution of the qubit's population inversion
and the entanglement behavior taking two qubits as an example. The
analytical methods provide unexpectedly accurate results in describing the
dynamics of the qubit in the present experimentally accessible coupling
regime, showing that the collapse-revival phenomena emerge, survive, and are
finally destroyed when the coupling strength increases beyond the
ultra-strong coupling regime. The inhomogeneous coupling system exhibits new
dynamics, which are different from homogeneous coupling case. The suggested
procedure applies readily to the multiple qubits system such as the GHZ
state entanglement evolution and quantum entanglement between a single
photon and superconducting qubits of particular experiment interest.
\end{abstract}

\maketitle

\section{Introduction}

The Jaynes-Cummings (JC) model with the rotating wave approximation (RWA),
first introduced in 1963 \cite{Jaynes}, is the simplest model that describes
the interaction between a two-level atom and a single mode quantized
radiation field. The RWA is applicable when the applied electromagnetic
field frequency $\omega _{c}$ is near resonance with the atom transition
frequency $\omega _{j}$ and the interaction between the atom and the
radiation field is weak. The reason is that the contribution of the
counter-rotating terms of the system is very small. Typical new era of
experiments witness the breakdown of the JC model in terms of both coupling
and detuning, including circuit QED experiments with superconducting qubits
coupled to LC and waveguide resonators \cite{Schuster, Hofheinz, Forn-Diaz,
Niemczyk} and Cooper-pair boxes or Josephson phase qubits coupled to
nanomechanical resonators \cite{Bouchiat, Wallraff, Nakamura, LaHaye,
Connell}. Each of these artificial atoms has an internal degree of freedom
(d.o.f.) that can be either up or down, creating a spin-1/2 system. These systems
generally allow coupling strengths up to $g_{j}/\omega _{c}\simeq 0.1$ in
the so-called ultrastrong coupling regime, or the qubit transition frequency
far-detuned from the field frequency \cite{Fedorov, Abdumalikov, Higgins,
Irish1, Amico, Irish2, Chen1, Werlang, Zheng, Zhang, Liu, He, He2, Wolf}. An
adiabatic approximation approach \cite{Irish1} was proposed to treat the
parameter regime outside the near-resonance and weak-coupling assumption of
RWA based on the displaced oscillator basis. In this basis the Hamiltonian
are truncated to a block-diagonal form and the blocks are solved
individually. The time evolution of the two-level-system occupation
probability with thermal and coherent state initial conditions for the
oscillator exhibits clear signals of collapse and revival.

The quantum Rabi model \cite{Rabi}, or JC model without the RWA, was
recently declared solved exactly in \cite{Braak, Solano}. By means of the
representation of bosonic operators in the Bargmann space Braak argued that
the regular spectrum of the Rabi model was given by the zeros of a
transcendental function, which is given as an infinite power series. Chen
et. al. mapped the model to a polynomial equation with a single variable in
terms of tunable extended bosonic coherent states \cite{Chen2}. They recover
Braak's exact solution in an alternative more physical way and point out
that both methods have one thing in common: the spectrum can not be obtained
without truncation in the power series \cite{Chen4}. Thus the Rabi model is
quasi-exactly solvable in the sense that only a finite part of the spectrum
can be obtained in closed form and the remaining part of the spectrum can
only be determined by numerical means \cite{Moroz1, Moroz2, Zhang2, Peng}.
The number of ca. 1350 calculable energy levels in each parity subspace are
obtained in double precision by an elementary stepping algorithm up to two
orders of magnitude higher than Braak's solution \cite{Moroz2}. Quantum
integrability is, according to Braak's criterion, equivalent to the
existence of quantum numbers that classify eigenstates uniquely. The Rabi
model is integrable because it has two d.o.f. and the eigenstates
can be uniquely labelled by two quantum numbers associated with the energy
and the parity, respectively. Moreover, examples of nonintegrable but
(quasi)exactly solvable system are given with broken parity symmetry or
vanished level splitting of an additional qubit \cite{Braak, Peng}. These
theoretical progress has renewed the interest in the Rabi and related
models. Analytical solutions of this model have brought clarity and
intuition to several important problems and experimental results in
contemporary quantum information. Furthermore, it is expected that the
experiments could reach the deep strong coupling regime \cite{Casanova,
Liberato} where the ratio of the coupling strength to the relevant
frequencies exceeds unity. Perturbative methods and the concept of Rabi
oscillations should be superseded by novel physics such as parity chains and
photon number wave packets.

To describe the collective behavior of multiple atomic dipoles interacting
with an electromagnetic field mode, the Dicke model \cite{Dicke} was
introduced where the Pauli operators are summed and transformed into a
bosonic operator. Though very successful in treating the system of an
alkaline atomic ensemble \cite{Baumann} in an optical microcavity with the
number of atoms over $10^5$, it does not apply well to the case of
multiqubit superconducting circuits with $N \le 10$. Theoretical studies for
a finite number $N$ of qubits in the system often employ the Tavis-Cummings
(TC) model \cite{Tavis} under RWA, where all the spins are grouped into a
total large spin. The TC model has received much attention and has been
involved in both experiments and theories \cite{Agarwal, Lopez, Ian, Chen3,
Chilingaryan, Braak1, Wilson, Wang, Lambert, Fink, DiCarlo}. Being exactly
solvable only when the coupling is homogeneous and when the eigenfrequencies
between the qubits and the photon mode are equal, it was recently \cite%
{Agarwal} extended beyond the RWA for quasi-degenerate qubits in the
parameter regime in which the frequencies of the qubits are much smaller
than the oscillator frequency and the coupling strength is allowed to be
ultrastrong. The case of inhomogeneous coupling is drastically different -
there is no such straightforward way to access the Hilbert space. The
extension to the inhomogeneous coupling system is limited to RWA \cite%
{Lopez, Ian} or numerical exact approach on the entanglement evolution of
two independent JC atoms \cite{Chen3}. It is worthwhile to
notice that more efforts are paid to the system composed of two nonidentical
qubits \cite{Chilingaryan}, or the $N = 3$ Dicke model which couples three
qubits to a single radiation mode and constitutes the simplest
quantum-optical system allowing for Greenberger-Horne-Zeilinger (GHZ) states
\cite{Braak1}.

We in this paper solve the TC model beyond the RWA with $N$ qubits coupled
to a single oscillator mode by comprehensively considering the recently
developed approaches in solving the JC model with extension to the case of
inhomogeneous coupling. A systematic truncated method is developed in finding
the exact wave functions of the model with $N$ discrete and one continuous d.o.f. 
In the basis of the displaced operators we construct the Hamiltonian by primitive 
building blocks and allow transitions between adjacent blocks in addition. We take 
mainly arbitrary two qubits interacting with a bose field as an example. The
analytical eigensolutions are derived in the zeroth-order and first-order
approximation to the exact wave function in deep strong coupling parameter
regimes. The procedure is easily extended to systems with more than two
qubits. We further examine the solvability and integrability of the system and
level crossing points in the energy spectrum of the same parity space  
are related to a hidden symmetry in the system. Then, starting
from any initial state of the system we are able to derive the time
evolution properties of the qubits by using a linear combination of the
analytical eigensolutions and tracing over the oscillator field. In other
words, some general techniques are applied to investigate the time evolution
of the rather complicated multiqubit-field system. Subsequently, we can
apply the approximated eigensolutions to study the dynamical evolution of
the entanglement between the two qubits as a fundamental consequence of
quantum mechanics and as a resource for communication and information
processing \cite{Bell, Wootters, Lee, Coffman, Lopez1}.

The paper is organized as follows. In Sec. II the analytical eigen solutions
of the TC model beyond the RWA is given after introducing the general
procedure of constructing the determinant of the secular equation. Sec. III
is devoted to the dynamical behaviors of the qubits, in which the analytical
eigensolutions are employed to approximately describe the time evolution of
the probability of finding both qubits in their initial state and the
population inversion when the quantum field is prepared initially in the
displaced coherent state. In Sec. IV we further study the entanglement
dynamics for two qubits starting from the Bell state and the field in a
coherent state. Finally, we make some discussions on the characteristics of
case of unequal coupling strengthes for the two qubits from the present
study in Sec. V.

\section{ Eigen solutions of the TC model beyond the RWA}

The system we consider here consists of $N$ qubits inhomogeneously
interacting with a single-mode bose field. It is described by the TC
Hamiltonian beyond the RWA (we set $\hbar =1 $) \cite{Tavis}%
\begin{equation}
H=\omega _{c}a^{\dagger }a+\sum\limits_{j=1}^{N}\left( -\frac{\omega _{j}}{2}%
\hat{\sigma} _{j}^{x}+g_{j}\left( a^{\dagger }+a\right) \hat{\sigma}%
_{j}^{z}\right),  \label{H}
\end{equation}%
where $a^{\dagger }$ ($a$) is the bosonic creation (annihilation) operator
of the single bosonic mode with frequency $\omega_c$, $\omega _{j}$ denotes
the energy splitting of $j$-th qubit described by Pauli matrices $\hat{\sigma%
}_{j}^{k}$ ($k=x,y,z$), and $g_{j}$ is the dipole-field coupling strength
between the qubit $j$ and the field. Here we have rotated the system around
the $y-$axis by an angle $\pi /4$ realized through a unitary transformation
\cite{Chen3} $V=\exp \left( \frac{i\pi }{4}\sum_{j}\hat{\sigma}%
_{j}^{y}\right) $. Basically the calculation here applies to arbitrary
non-identical two-level atoms and non-uniform coupling strengths $g_j$ in
any form. We also note that the Hamiltonian (\ref{H}) conserves the global
parity operator defined as $\Pi =\prod\limits_{j=1}^{N}\hat{\sigma}
_{j}^{x}\exp (i\pi a^{\dagger }a)$, i.e. $\left[ H,\Pi \right] =0$. In the
following we use the parity operator to decompose the Hilbert space into
even and odd subspaces.

For the convenience of description, we take $N=2$ as an illustrative
example. Denote the upper and lower eigenstates of $\hat{\sigma}_{j}^{z}$ as
$\left\vert 1\right\rangle _{j}$ and $\left\vert 0\right\rangle _{j}$
respectively. Formally we treat qubit 2 as a new member to the Rabi model
\cite{Rabi,Braak} and unfold the dimension of the space from 2 to 4. In the
basis of product space of the two qubits, i.e. $\left\vert 11\right\rangle
=\left\vert 1\right\rangle _{2}\otimes \left\vert 1\right\rangle _{1},$ $%
\left\vert 10\right\rangle =\left\vert 1\right\rangle _{2}\otimes \left\vert
0\right\rangle _{1},$ $\left\vert 01\right\rangle =\left\vert 0\right\rangle
_{2}\otimes \left\vert 1\right\rangle _{1}$, and $\left\vert 00\right\rangle
=\left\vert 0\right\rangle _{2}\otimes \left\vert 0\right\rangle _{1}$, we
may write the Hamiltonian into a matrix form
\begin{widetext}
\begin{eqnarray}
H=\left(
\begin{array}{cccc}
\omega_{c} \left( a^{\dagger }a+\beta _{1}\left( a^{\dagger }+a\right) \right) &
-\frac{\omega _{1}}{2} & -\frac{\omega _{2}}{2} & 0 \\
-\frac{\omega _{1}}{2} & \omega_{c} \left( a^{\dagger }a+\beta _{2}\left(
a^{\dagger }+a\right) \right) & 0 & -\frac{\omega _{2}}{2} \\
-\frac{\omega _{2}}{2} & 0 & \omega_{c} \left( a^{\dagger }a+\beta _{3}\left(
a^{\dagger }+a\right) \right) & -\frac{\omega _{1}}{2} \\
0 & -\frac{\omega _{2}}{2} & -\frac{\omega _{1}}{2} & \omega_{c} \left(
a^{\dagger }a+\beta _{4}\left( a^{\dagger }+a\right) \right)%
\end{array}%
\right)  \label{4}
\end{eqnarray}%
\end{widetext}We notice that the original $2\times 2$ Hamiltonian \cite{Liu}
as a primitive block is shifted along the diagonal line with $g_{1}$ and $%
g_{2}$ being recombined into 4 coupling parameters $\beta _{i}$ with
relations $\beta _{1}=-\beta _{4}=\left( g_{2}+g_{1}\right) /\omega _{c}$
and $\beta _{2}=-\beta _{3}=\left( g_{2}-g_{1}\right) /\omega _{c}$, while
the transition frequency $\omega _{2}$ always appears in the off-diagonal
blocks of the matrix. Similar procedure can be applied when we add one more
qubit to the system. The basis of $N$ qubits is $2^{N}$ dimension and we
have $2^{(N-1)}$ independent dimensionless coupling parameters $\beta _{i}$.
A special case is the system of two identical qubits ($\omega _{1}=\omega
_{2}$) coupling with a common oscillator mode, which has been extended
beyond the RWA in Ref. \cite{Agarwal}. By assuming that the coupling
parameters are larger than the transition frequencies, i.e. in the
deep-strong-coupling regime, $g_{1},g_{2}\gg \omega _{1},\omega _{2}$, the
eigenvalues have been calculated up to the second-order perturbation
correction \cite{Chilingaryan}. The main results for these limiting cases
may be readily reproduced from the method in this paper.

Let us first introduce the displacement operators \cite{Schweber,Irish1} $%
\hat{D}\left( \beta _{i}\right) =\exp \left[ \beta _{i}\left( a^{\dagger
}-a\right) \right] $, which will translate the field operators $a^{\dagger }$
and $a$ by a distance $\beta_i$ and give rise to $A_{i}^{\dagger } =\hat{D}%
^{\dagger }\left( \beta _{i}\right) a^{\dagger }\hat{D}\left( \beta
_{i}\right) =a^{\dagger }+\beta _{i}$ and $A_{i}=\hat{D}^{\dagger }\left(
\beta _{i}\right) a\hat{D}\left( \beta _{i}\right) =a+\beta _{i}$, and will
transform the number state $\left\vert n\right\rangle$ defined as $%
a^{\dagger }a\left\vert n\right\rangle =n\left\vert n\right\rangle $ into
the so-called displaced Fock number state defined as $A_{i}^{\dagger
}A_{i}\left\vert n\right\rangle _{A_{i}} =n\left\vert n\right\rangle
_{A_{i}} $, i.e. $\hat{D}^{\dagger }\left( \beta _{i}\right) \left\vert
n\right\rangle =\left\vert n\right\rangle _{A_{i}}$. The states $\left\vert
n\right\rangle _{A_{i}}$ are orthogonal for the same index $i$ and
non-orthogonal for different subspaces $i$ and $j$. The lack of
orthogonality between states with different displacements leads to the
unusual results in the dynamics of population inversion and entanglement
which will be shown later in this work.

The diagonal elements of the Hamiltonian are in this way reconstructed as $%
\omega _{c}\left( A_{i}^{\dagger }A_{i}-\beta _{i}^{2}\right) $ with the
off-diagonal $\omega _{j}$ unchanged. The Hilbert space of the diagonal
Hamiltonian is now of the form of a combination of qubit basis and displaced
oscillator basis, e.g. $\left\vert 111...\right\rangle \left\vert
n\right\rangle _{A_{1}}$, which can be taken as a starting point to expand
the eigenfunction of the total Hamiltonian $H$. For two-qubit system we
suppose that
\begin{eqnarray}
\left\vert \psi \right\rangle &=&\sum_{n=0}^{\infty }\left( d_{1n}\left\vert
11\right\rangle \left\vert n\right\rangle _{A_{1}}+d_{2n}\left\vert
10\right\rangle \left\vert n\right\rangle _{A_{2}}\right.  \notag \\
&&\left. +d_{3n}\left\vert 01\right\rangle \left\vert n\right\rangle
_{A_{3}}+d_{4n}\left\vert 00\right\rangle \left\vert n\right\rangle
_{A_{4}}\right)  \label{eigenfunction}
\end{eqnarray}%
which in addition should be the eigenfunction of the parity operator $\Pi $,
i.e. $\Pi \left\vert \psi \right\rangle =\kappa \left\vert \psi
\right\rangle $ with $\kappa =+,-$ for even and odd parity respectively.
Actually we find that the $\sigma $'s operators in $\Pi $ transform the
qubit basis from $\left\vert 11\right\rangle $ ($\left\vert 10\right\rangle $%
) to $\left\vert 00\right\rangle $ ($\left\vert 01\right\rangle $) and vice
versa, while the field operators set up links between the displaced Fock
state $\left\vert n\right\rangle _{A_{1}}$ ($\left\vert n\right\rangle
_{A_{2}}$) and its symmetric counterpart $\left\vert n\right\rangle _{A_{4}}$
($\left\vert n\right\rangle _{A_{3}}$). Thus the coefficients are related to
each other through $d_{4n}=\kappa \left( -1\right) ^{n}d_{1n}$ and $%
d_{3n}=\kappa \left( -1\right) ^{n}d_{2n}$. This symmetry separates the
state space into two different invariant subspaces can be labeled by the
eigenvalues of the operation $\Pi $ with $\kappa =+,-.$ In accordance with
the Schr\"{o}dinger equation we find that the number of equations is reduced
by half
\begin{eqnarray}
\varepsilon _{1m}d_{1m}+\sum_{n=0}^{\infty }\Omega _{mn}^{\kappa }d_{2n}
&=&Ed_{1m},  \label{Schrdinger equation1} \\
\varepsilon _{2m}d_{2m}+\sum_{n=0}^{\infty }W_{mn}^{\kappa }d_{1n}
&=&Ed_{2m},  \label{Schrdinger equation2}
\end{eqnarray}%
where $\varepsilon _{im}=\left( m-\beta _{i}^{2}\right) \omega _{c}$ and the
off-diagonal terms describe the transitions between states belonging to
different displaced Fock spaces
\begin{eqnarray*}
\Omega _{mn}^{\kappa } &=&-\frac{\omega _{1}}{2}\left[ _{A_{1}}\left\langle
m|n\right\rangle _{A_{2}}\right] -\kappa \left( -1\right) ^{n}\frac{\omega
_{2}}{2}\left[ _{A_{1}}\left\langle m|n\right\rangle _{A_{3}}\right] , \\
W_{mn}^{\kappa } &=&-\frac{\omega _{1}}{2}\left[ _{A_{2}}\left\langle
m|n\right\rangle _{A_{1}}\right] -\kappa \left( -1\right) ^{n}\frac{\omega
_{2}}{2}\left[ _{A_{2}}\left\langle m|n\right\rangle _{A_{4}}\right] .
\end{eqnarray*}%
Clearly the symmetry of the coefficients $d_{im}$ reduces the number of
equations by half. This effectively folds the already expanded Hilbert space
to a diagonal block, in which the anti-diagonal elements are occupied by the
newly added $\omega _{N}$ together with the parity $\kappa $ as in the
expressions for $\Omega $ and $W$ above. The nonzero off-diagonal elements
originates from the non-orthogonality of displaced Fock states \cite{Liu}
\begin{equation}
_{A_{i}}\left\langle m|n\right\rangle _{A_{j}}=e^{\frac{-\beta _{ij}^{2}}{2}%
}\sum_{l}^{min(m,n)}\frac{(-1)^{n-l}\sqrt{m!n!}}{l!(n-l)!(m-l)!}\beta
_{ij}^{m+n-2l}  \label{Xmn}
\end{equation}%
with $\beta _{ij}=\beta _{i}-\beta _{j}$. This implies that the interchange
of $i$ and $j$, that of $m$ and $n$, or the inversion of $\beta _{i}$ to $%
-\beta _{i}$ will introduce a factor $(-1)^{m+n}$ in eq. (\ref{Xmn}). Then
in terms of expressions about $\Omega _{mn}^{\kappa }$ and $W_{mn}^{\kappa }$%
, we find the symmetry relation
\begin{equation}
\Omega _{mn}^{\kappa }=W_{nm}^{\kappa },  \label{OmegaW}
\end{equation}%
which turns the Hamiltonian in the Hilbert space into a real symmetric
matrix which assures that all coefficients $d_{im}$ are real. This allows us
keep only $\Omega $ in the rest of the paper.

\begin{figure}[t]
\includegraphics[width=0.45\textwidth]{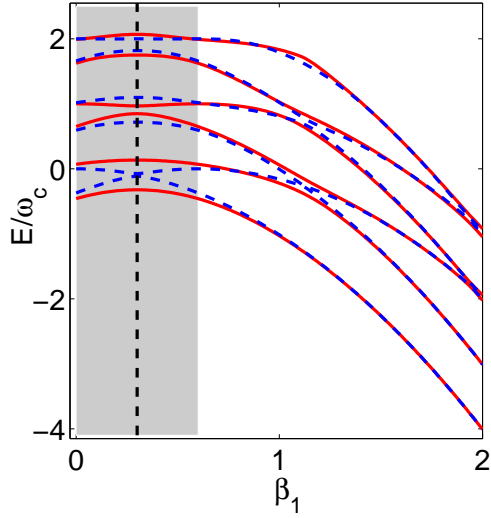}
\caption{(Color Online) The ED numerical solution of the energy levels as a
function of coupling strength $\protect\beta_{1}$ in the even (red solid
lines) or odd (blue dashed lines) party subspaces for $\protect\omega _{1}=%
\protect\omega _{2}=0.25\protect\omega _{c}$, $g_{1}=0.3\protect\omega _{c}$%
. In the shadowed area the energy levels are symmetric about the vertical
line $\protect\beta_1=0.3$, i.e. $g_2=0$.}
\label{Fig1}
\end{figure}

The $2^{N-1}$ equations for $N$ qubits take similar forms as eqs. (\ref%
{Schrdinger equation1}) and (\ref{Schrdinger equation2}) with the diagonal
terms $\varepsilon _{im}d_{im}$ and in each equation off-diagonal terms indicate 
the transitions between Fock
states displaced in different directions and distances. It is easy to show
that the eigenvalue spectrum of the $N$ qubits system is unaltered when any
of the coupling strengths changes its sign, e.g. $g_{1}\rightarrow -g_{1}$
or $g_{2}\rightarrow -g_{2}$ so that it suffices to discuss the energy
spectrum for positive values of $g_{i}$. The equations (\ref{Schrdinger
equation1}) and (\ref{Schrdinger equation2}) are solved by means of exact
diagonalization (ED) and the energy spectrum are shown in Fig. \ref{Fig1} as
functions of $\beta _{1}$ for $g_{1}=0.3\omega _{c}$ and $\omega _{1}=\omega
_{2}=0.25\omega _{c}$. As mentioned earlier, the power series (\ref%
{eigenfunction}) has to be truncated in order to obtain the spectrum. Here
we set the truncation number $n_{tr}=48$ such that the calculation is done
in a closed subspace $\left\vert n\right\rangle _{A_{i}}$ with $\left(
n=0,1,2\cdots ,n_{tr}\right) $ and the off-digonal elements $\Omega _{mn}$
for $m,n>n_{tr}$ are less than $10^{-6}$. The lowest 6 levels are
illustrated in Fig. \ref{Fig1} for even and odd parities respectively and we
find similar to the single qubit case, strong coupling strength tends to
lower the eigen-energies of the system and level crossing occurs for
different parities, while levels with the same parity prefers to avoid this.
Furthermore, the energy spectrum is symmetric about the vertical dashed line
$\beta _{1}=0.3$ in the shadowed area (correspondingly $g_{2}/\omega _{c}$
takes the value between $-0.3$ and $0.3$).

In any case, the process of getting analytically the eigenvalues and
eigenvectors for the $N$-qubit system is very difficult, if not impossible,
so approximation techniques have to be employed. Here we illustrate how to
identify the building blocks of the determinant by taking the two-qubit
system as an example, and the procedure of solving the eigen-equations up to
any order of approximations to the exact result of the wave function. The
condition for the existence of non-trivial solutions of $d_{im}$ is the
secular equation described by the determinant
\begin{widetext}
\begin{eqnarray}
\left\vert
\begin{array}{cccccc}
\!\cdots\! & \vdots & \vdots & \vdots & \vdots & \!\cdots\! \\
\!\cdots\! & \!\varepsilon _{1m}\!\!-\!E & \Omega _{mm}^{\kappa } & 0 & \Omega _{m\left(
m+1\right) }^{\kappa } & \cdots \\
\!\cdots\! & \Omega_{mm}^{\kappa } & \!\varepsilon _{2m}\!\!-\!E & \Omega_{\left( m+1\right)m }^{\kappa }
& 0 & \!\cdots\! \\
\!\cdots\! & 0 & \Omega _{\left( m+1\right) m}^{\kappa } & \!\varepsilon _{1\left(
m+1\right) }\!\!-\!E & \Omega _{\left( m+1\right)\left( m+1\right) }^{\kappa } &
\!\cdots\! \\
\!\cdots\! & \Omega_{m\left( m+1\right) }^{\kappa } & 0 & \Omega_{\left( m+1\right)\left( m+1\right) }^{\kappa } & \!\varepsilon _{2\left( m+1\right) }\!\!-\!E & \!\cdots \!\\
\!\cdots\! & \vdots & \vdots & \vdots & \vdots & \!\cdots\!%
\end{array}%
\right\vert =0.  \label{Det}
\end{eqnarray}%
\end{widetext}
We see that in the two-qubit case the primitive building block of the
determinant is $2 \times 2$ and in the first-order approximation two blocks $%
m$ and $m+1$ are involved, the transitions between which are determined by
the off-diagonal elements $\Omega_{m(m+1)}$ and $\Omega_{(m+1)m}$. The
primitive building block for $N$ qubits is $2^{N-1}\times 2^{N-1}$, while
higher-order approximation involves more identical blocks with $m$ increased
with step $1$ and those off-diagonal $\Omega $ terms induce transition
between blocks with different $m$.

\subsection{Zeroth-order approximation}

The key property of the inhomogeneously coupled $N$-qubit system is
exhibited by the zeroth-order approximation of equation (\ref{Det}), which
neglects all transitions between states with different $m$. This is often
called the adiabatic approximation. Similar approximated solution to the JC
model without the RWA is shown to be valid when the transition frequency of
the qubit $\omega _{0}$ is much smaller than the frequency of the
single-mode bose field $\omega _{c}$ and it is very efficient for coupling
strengths $g$ up to or larger than the oscillator frequency \cite{Irish1}.
For the $N$-qubit system we first consider the zeroth-order approximation
and truncate the determinant Eq. (\ref{Det}) to the lowest order. This
leaves us a block with the same index $m$, the diagonal terms of which read
as $\varepsilon _{im}-E$ with $i=1,2...2^{N-1}$ and the off-diagonal
transition terms are those coefficients $_{A_{i}}\left\langle
m|m\right\rangle _{A_{j}}$. For two-qubit system the zeroth-order
approximation for the determinant takes the following block form
\begin{equation}
\left\vert
\begin{array}{cc}
\varepsilon _{1m}-E & \Omega _{mm}^{\kappa } \\
\Omega _{mm}^{\kappa } & \varepsilon _{2m}-E%
\end{array}%
\right\vert =0.  \label{hanglieshi2}
\end{equation}%
For convenience, we denote $\Omega _{mm}^{\kappa }$ as $\Omega _{m}^{\kappa
} $. Consequently the solutions for the eigenergies are
\begin{equation}
E_{m}^{\kappa \pm }=m\omega _{c}-\left( \beta _{1}^{2}+\beta _{2}^{2}\right)
\omega _{c}/2\pm \theta _{m}^{\kappa }  \label{zero app}
\end{equation}%
with $\theta _{m}^{\kappa }=\sqrt{\left( \Omega _{m}^{\kappa }\right)
^{2}+\omega _{c}^{2}\left( \beta _{1}^{2}-\beta _{2}^{2}\right) ^{2}/4}$.
Based on the symmetry of the coefficients $d_{im}$, the eigenstates of the
system that satisfy the orthogonality and completeness conditions have the
form
\begin{equation}
\left\vert \psi _{m}^{\kappa \pm }\right\rangle =\frac{1}{\sqrt{2}}\left(
\begin{array}{c}
d_{1m}^{\kappa \pm } \\
d_{2m}^{\kappa \pm } \\
(-1)^{m}\kappa d_{2m}^{\kappa \pm } \\
(-1)^{m}\kappa d_{1m}^{\kappa \pm }%
\end{array}%
\right) ,  \label{function}
\end{equation}%
where
\begin{equation*}
d_{1m}^{\kappa \pm }=\xi _{m}^{\kappa \pm }\sqrt{\frac{1}{1+\left( \xi
_{m}^{\kappa \pm }\right) ^{2}}},d_{2m}^{\kappa \pm }=-\sqrt{\frac{1}{%
1+\left( \xi _{m}^{\kappa \pm }\right) ^{2}}},
\end{equation*}%
with $\xi _{m}^{\kappa \pm }=\Omega _{m}^{\kappa }/\left( \left( \beta
_{2}^{2}-\beta _{1}^{2}\right) \omega _{c}/2\mp \theta _{m}^{\kappa }\right)
$.

A special case is the system with two completely identical qubits
homogeneously coupled to a bose field, which means that $\beta _{2}=0$ and $%
\omega _{1}=\omega _{2}$. This simplifies the transition frequency as $%
\Omega _{m}^{\kappa }\sim \left( 1+\kappa (-1)^{m}\right) $, i.e. the parity
of the Hamiltonian ($\kappa $) and that of the displaced Fock space ($m$)
together decide whether $\Omega _{m}^{\kappa }$ is zero or not. When $\Omega
_{m}^{\kappa }=0$ we have the eigenenergies
\begin{equation}
E_{m}^{\kappa +}=m\omega _{c},E_{m}^{\kappa -}=\left( m-\beta
_{1}^{2}\right) \omega _{c},  \label{zero2}
\end{equation}%
and the corresponding eigenfunctions $\left\vert \psi _{m}^{\kappa
+}\right\rangle =\left\vert \psi _{1}\right\rangle $, $\left\vert \psi
_{m}^{\kappa -}\right\rangle =\left\vert \psi _{2}\right\rangle ,$ with
\begin{equation}
\left\vert \psi _{1}\right\rangle =\frac{1}{\sqrt{2}}\left(
\begin{array}{c}
0 \\
1 \\
-1 \\
0%
\end{array}%
\right) ,\left\vert \psi _{2}\right\rangle =\frac{1}{\sqrt{2}}\left(
\begin{array}{c}
1 \\
0 \\
0 \\
-1%
\end{array}%
\right) .  \label{singlet}
\end{equation}%
For nonzero $\Omega _{m}^{\kappa }$ we assume $\left\vert \Omega
_{m}^{\kappa }\right\vert /\omega _{c}\gg \beta _{1}^{2}$ as in Ref. \cite%
{Agarwal}, the eigenergies can be expressed as
\begin{equation}
E_{m}^{\kappa +}=m\omega _{c}+\Omega _{m}^{\kappa },E_{m}^{\kappa -}=m\omega
_{c}-\Omega _{m}^{\kappa },  \label{zero app2}
\end{equation}%
with eigenfunctions $\left\vert \psi _{m}^{\kappa +}\right\rangle
=\left\vert \psi _{3}\right\rangle $, $\left\vert \psi _{m}^{\kappa
-}\right\rangle =\left\vert \psi _{4}\right\rangle ,$ with%
\begin{equation}
\left\vert \psi _{3}\right\rangle =\frac{1}{2}\left(
\begin{array}{c}
1 \\
1 \\
1 \\
1%
\end{array}%
\right) ,\left\vert \psi _{4}\right\rangle =\frac{1}{2}\left(
\begin{array}{c}
1 \\
-1 \\
-1 \\
1%
\end{array}%
\right) .  \label{zero1}
\end{equation}

\begin{figure}[t]
\includegraphics[width=0.45\textwidth]{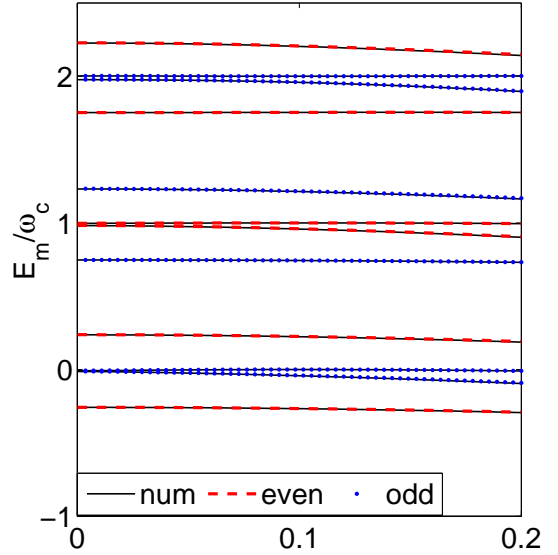}
\caption{(Color Online) The zeroth-order approximation of the energy levels
with even and odd parities as a function of coupling strength $g_{2}/\protect%
\omega _{c}$ for $\protect\omega _{1}=\protect\omega _{2}=0.25\protect\omega %
_{c}$ and $g_{1}$ = $0.1\protect\omega _{c}$, compared to the ED numerical
ones. }
\label{Fig2}
\end{figure}

We note that in writing down these four eigenstates of $H$ the qubit basis
are chosen as the uncoupled representation of spin operators $\hat{\sigma}%
_{1,2}^{z}$ in order to solve the qubits system with different frequencies
and coupling strength. Three of states are alternatively \cite{Agarwal}
expanded in terms of the triplet states of the total spin component $S_{x}$
of the two identical qubits in the case of homogeneous coupling. The spin
singlet state is exactly $\psi _{1}$ in (\ref{singlet}) which is itself the
eigenstate of the Hamiltonian. By including the singlet state into the
eigenvectors, we are enable to treat the dynamics of any initial state of
the system. States such as $\left\vert 10\right\rangle $ or $\left\vert
01\right\rangle $, i.e. when the two qubits are respectively put in the
upper and lower eigenstates of their $\hat{\sigma}^{z}$s, are out of reach
in the triplet manifold, which will be seen this in the next section. In
Fig. \ref{Fig2}, we can see that the analytical results in the zeroth-order
approximation already agree with the exact solutions very well in the
ultra-strong coupling case $g_{2}\sim 0.2\omega _{c}$ $(g_{1}=0.1\omega
_{c}) $, where each four eigenenergies with the same index $m$ bundle into a
group corresponding to the four qubit basis $\left\vert 00\right\rangle
,\left\vert 01\right\rangle ,\left\vert 10\right\rangle ,\left\vert
11\right\rangle $ in the absence of coupling. In each group, the parity of
each eigenstate is fixed with the lowest level being always even parity. For
even $m$, two odd parity levels are held between two even parity ones, or
vice versa for odd $m$. This can be compared to the single-qubit case where
the energy levels are arranged as (even, odd), (odd, even), (even, odd),
etc. because the qubit basis are instead $\left\vert 0\right\rangle
,\left\vert 1\right\rangle $ and adding one photon will alter the parity.

\subsection{First-Order Approximation}

Next we consider the first-order approximation for the determinant (\ref{Det}%
) and permit transitions between blocks $m$ and $m+1$ \cite{Liu}. In the
case of two-qubit system we consequently make a cut-off in Eq. (\ref{Det})
such that
\begin{equation}
\left\vert
\begin{array}{cccc}
\varepsilon _{1m}\!-\!E & \Omega _{m}^{\kappa } & 0 & \Omega _{m\left(
m\!+\!1\right) }^{\kappa } \\
\Omega _{m}^{\kappa } & \varepsilon _{2m}\!-\!E & \Omega _{\left(
m\!+\!1\right) m}^{\kappa } & 0 \\
0 & \Omega _{\left( m\!+\!1\right) m}^{\kappa } & \varepsilon _{1(m+1)}-E &
\Omega _{m+1}^{\kappa } \\
\Omega _{m\left( m\!+\!1\right) }^{\kappa } & 0 & \Omega _{m+1}^{\kappa } &
\varepsilon _{2(m+1)}\!-\!E%
\end{array}%
\right\vert =\!0.  \label{hanglieshi}
\end{equation}%
The equation can be solved analytically because it leads to a quartic
equation in the form of $E^{4}+bE^{3}+cE^{2}+dE+e=0$. The coefficients in
the quartic equation are assigned for parity $\kappa $ and index $m$ and
expressed as (for simplicity we drop the superscript and subscript)
\begin{eqnarray*}
b &=&-\varepsilon _{1m}-\varepsilon _{2m}-\varepsilon _{1(m+1)}-\varepsilon
_{2(m+1)}, \\
c &=&\left( \varepsilon _{1m}+\varepsilon _{2m}\right) \left( \varepsilon
_{1(m+1)}+\varepsilon _{2(m+1)}\right) \\
&+&\varepsilon _{1m}\varepsilon _{2m}+\varepsilon _{1\left( m+1\right)
}\varepsilon _{2\left( m+1\right) }-(\Omega _{m}^{\kappa })^{2} \\
&-&(\Omega _{m+1}^{\kappa })^{2}-(\Omega _{m(m+1)}^{\kappa })^{2}-(\Omega
_{\left( m\!+\!1\right) m}^{\kappa })^{2}, \\
d &=&\left( \varepsilon _{1m}\!+\!\varepsilon _{2\left( m\!+\!1\right)
}\right) (\Omega _{\left( m\!+\!1\right) m}^{\kappa })^{2} \\
&+&\left( \varepsilon _{1\left( m\!+\!1\right) }\!+\!\varepsilon
_{2m}\right) (\Omega _{m(m+1)}^{\kappa })^{2} \\
&+&((\Omega _{m}^{\kappa })^{2}-\varepsilon _{1m}\varepsilon
_{2m})(\varepsilon _{1(m+1)}+\varepsilon _{2(m+1)}) \\
&+&(\varepsilon _{1m}+\varepsilon _{2m})((\Omega _{m+1}^{\kappa
})^{2}-\varepsilon _{1(m+1)}\varepsilon _{2(m+1)}),
\end{eqnarray*}%
and%
\begin{equation*}
e=\left\vert
\begin{array}{cccc}
\varepsilon _{1m} & \Omega _{m}^{\kappa } & 0 & \Omega _{m(m+1)}^{\kappa }
\\
\Omega _{m}^{\kappa } & \varepsilon _{2m} & \Omega _{\left( m\!+\!1\right)
m}^{\kappa } & 0 \\
0 & \Omega _{\left( m\!+\!1\right) m}^{\kappa } & \varepsilon _{1(m+1)} &
\Omega _{m+1}^{\kappa } \\
\Omega _{m(m+1)}^{\kappa } & 0 & \Omega _{m+1}^{\kappa } & \varepsilon
_{2(m+1)}%
\end{array}%
\right\vert .
\end{equation*}%
For each given parity $\kappa $ and block index $m$ we get in general four
analytical solutions for the eigenenergies of Hamiltonian $H$
\begin{equation}
E_{m}^{\kappa }=-\frac{b}{4}\pm _{\gamma }Y\pm _{s}\frac{1}{2}\sqrt{-\left(
4Y^{2}+2p\pm _{\gamma }\frac{q}{Y}\right) },  \label{first app}
\end{equation}%
where the two occurrences of $\pm _{\gamma }$ must denote the same sign,
while $\pm _{s}$ can take its sign independently. The notations relating to
the coefficients of the quartic equation are defined as $p=c-3b^{2}/8$, $%
q=\left( b^{3}-4bc+8d\right) /8$, $\Delta _{0}=c^{2}-3bd+12e$, $\Delta
_{1}=2c^{3}-9bcd+27b^{2}e+27d^{2}-72ce$, and
\begin{eqnarray*}
Y &=&\frac{1}{2}\sqrt{-\frac{2p}{3}+\frac{Q+\Delta _{0}/Q}{3}}, \\
Q &=&\sqrt[3]{\frac{\Delta _{1}+\sqrt{\Delta _{1}^{2}-4\Delta _{0}^{3}}}{2}}.
\end{eqnarray*}

The first-order approximation improves the analytical results applicable
even in the deep coupling region $g>1$ as shown in Fig. 3, where we set $%
g_{1}=0.3\omega _{c}$ and $\omega_{1}=\omega_{2}=0.25 \omega_c$. According
to the assumption of the wavefunction (\ref{eigenfunction}), the dimension
of the Hilbert space depends on the truncation of the displaced Fock number
state as $4\left( n_{tr}+1\right)$. Correspondingly, for each $m$ we only
have four genuine solutions. The zeroth-order approximation permits exactly
four eigensolutions for each $m$ as shown above, while in the first-order
approximation one has eight eigensolutions for each combination $\left(
m,m+1\right) $ including four even parity and four odd parity solutions.

\begin{figure}[t]
\includegraphics[width=0.45\textwidth]{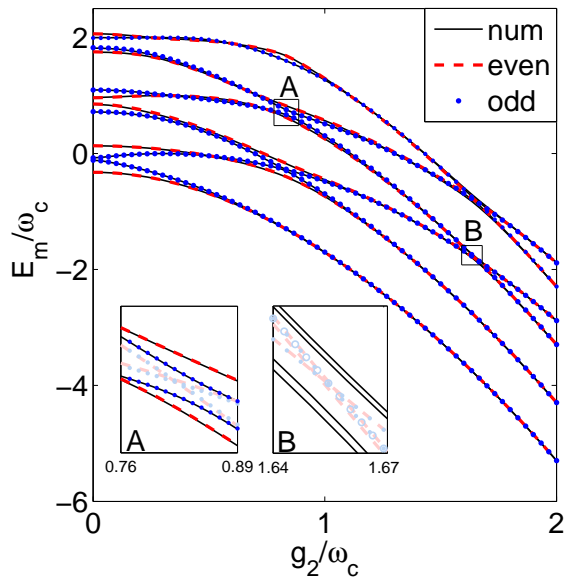}
\caption{(Color Online) The first-order approximation solution of the energy
levels as a function of coupling strength $g_{2}/\protect\omega_{c}$ with $%
\protect\omega _{1}=\protect\omega _{2}=0.25\protect\omega _{c}$ and $g_{1}$=%
$0.3\protect\omega_{c}$, compared to the ED numerical ones. Insets show two
typical level crossing points where the analytical results match the
numerical ones (left) for $\protect\beta_1 \sim 1$, or, on the other hand,
where they mismatch (right) for even stronger coupling $\protect\beta_1 \sim
2$. The pseudo-solutions in light-colored lines should be ruled out so that
the analytical first-order results agree with the numerical ones perfectly. }
\label{Fig3}
\end{figure}

Obviously the analytical results of first-order approximation show energy
level crossing between states with the same parity or different parities.
The level crossing of states with different parities are accidental and we
will focus on those with the same parity. On the uncertainty whether or not
energy levels can cross by changing parameters we give an argument in terms
of von Neumann and Wigner non-crossing theorem: For a real symmetric
(hermitian) matrix, we need to tune two (three) parameters to get a level
crossing \cite{Neumann}. In the case of two-qubit system, the Hamiltonian in
the Hilbert space is a real symmetric matrix. To determine the crossing or
anti-crossing we calculate the fidelity between appropriate states before
and after the crossing which is a measure of the "closeness" of two quantum
states and defined as $F=\left\vert \left\langle \phi |\varphi \right\rangle
\right\vert ^{2}$ for pure states \cite{Fidelity, Zanardi, Quan, Gu, ShuChen}%
. We remark that (1) for fixed $m$ the energy levels can never cross by
changing one parameter $g_{2}$; (2) However, for different $m$ the energy
level crossings may occur at some isolated points because two parameters $m$
and $g_{2}$ are tuned. In the left (A) inset of Fig. \ref{Fig3}, we zoom out
a particular anti-crossing point of exact results of energy levels. Four
analytical levels (deep-colored lines) in the first-order approximation
match the exact results which avoid crossing, while the rest (light-colored
lines) four curves mismatch and cross each other which should be discarded.

Taking the above aspects into consideration, for each invariant parity
subspace we must rule out half solutions of the first-order approximation by
keeping only solutions (\ref{first app}) with $\pm _{\gamma }$ opposite to $%
\pm _{s}$ for each combination $\left( m,m+1\right) $ because each $m$ has
been used twice in our calculation. An exception is the combination $(0,1)$,
in which case we only drop the solution with both signs positive. In this
way the pseudo solutions are removed and the analytical eigenvalues for the
two-qubit system agree perfectly with the exact results in the deep coupling
limit $g_{2}\sim 1.5\omega _{c}$ $(g_{1}=0.3\omega _{c})$ in Fig. \ref{Fig3}%
. The genuine solutions are two-fold degenerate in the deep strong coupling
regime $g_{2}>1$. This degeneracy has been found in the quantum Rabi model
for two qubits \cite{Chilingaryan}, though for both coupling parameters $%
g_{1},g_{2}$ larger than the transition frequencies of the qubits.

Moreover, in the right inset (B) of in Fig. \ref{Fig3}, as $g_{2}$ increases
across the next level anti-crossing point all the solutions of the
first-order approximation mismatch the exact results which suggests that
higher order approximation is needed. Our procedure applies readily to the
second-order approximation, in which case one has 12 solutions for each
combination of three blocks $(m,m+1,m+2)$ and the degeneracy grows rapidly.
In this way the approximated solutions will match the exact results in the
right inset of Fig. \ref{Fig3}. In particular, we notice that the
transitions would be suppressed if the off-diagonal matrix elements are much
smaller than the energy difference between the states belonging to different
blocks. Specifically, all elements $\left\vert \Omega _{nm}^{\kappa
}\right\vert $, $\left\vert \Omega _{mn}^{\kappa }\right\vert $ with $n\neq
m $ are set to zero in the zeroth-order approximation, while those with $%
n\neq m,m\pm 1$ are negligible in the first-order approximation.

\subsection{Solvability and Integrability}

In quantum mechanics there exist potentials for which it is possible to find
a finite number of exact eigenvalues and associated eigenfunctions in the
closed form. These systems are said to be quasi exactly solvable. The Rabi
model is a typical example distinguished by the fact that part of its
eigenvalues and corresponding eigenfunctions can be determined algebraically
for special values of the energy splitting of the qubit $\omega $ and the
coupling strength $g$ \cite{Moroz1,Moroz2,Zhang2}. Known as Judd's isolated
solutions \cite{Judd}, these exceptional spectrum with energy eigenvalues $%
E=n-g^{2}/\omega _{c}^{2}$ constitute the exact part of Rabi model and
doubly degenerate with respect to parity.

Here we show that the two-qubit TC model provides another example of
quasi-exactly solvable models, i.e part exact spectrum of the model can be
obtained in some special parameter region. First of all, in the homogeneous
coupling case $g_{1}=g_{2}$, there always exists a constant solution $%
E=\omega _{c}$ corresponding to either the even parity eigenstates
\begin{equation*}
\left\vert \psi _{e}\right\rangle =\left( q_{e}\left( \left\vert
01\right\rangle -\left\vert 10\right\rangle \right) |1\rangle +\left\vert
11\right\rangle |0\rangle \right) /\sqrt{2q_{e}^{2}+1},
\end{equation*}%
for the symmetric detunings with $\omega _{1}+\omega _{2}=2\omega _{c}$
(suppose $\omega _{1}>\omega _{2}$), or the odd parity eigenstates
\begin{equation*}
\left\vert \psi _{o}\right\rangle =\left( q_{o}\left( \left\vert
00\right\rangle -\left\vert 11\right\rangle \right) |1\rangle +\left\vert
01\right\rangle |0\rangle \right) /\sqrt{2q_{o}^{2}+1},
\end{equation*}%
for the asymmetric detunings with $\omega_{1}-\omega _{2}=2\omega _{c}$ with
$q_{e,o}=2g/\left( \omega _{1}\mp \omega _{2}\right) $. Secondly, for two
completely identical qubits homogeneously coupled to the bose field, i.e. $%
g_{1}=g_{2}$ and $\omega_{1}=\omega _{2}$, it is easy to prove the state $%
\left\vert \psi_{1}\right\rangle =\left( \left\vert 10\right\rangle
-\left\vert 01\right\rangle \right) |m\rangle /\sqrt{2}$ in (\ref{singlet})
for any $m$ is exactly the eigenstate of $H$ with eigenvalue $E_{m}=m\omega
_{c}$. The state has even(odd) parity for odd(even) $m$. 
Very recently, an alternative form of analytical solution is given to the quantum 
Rabi models with two identical qubits in
a similar way, however, essentially different from the Juddian solutions
with doubly degenerate eigenvalues in the one-qubit quantum Rabi model \cite%
{HuiWang}. In short, for the TC model with two qubits a finite part of the
spectrum can be obtained in closed form and the remaining part of the
spectrum is only numerically accessible. So we conclude that the TC model
with two qubits is quasi exactly solvable.

\begin{figure}[t]
\includegraphics[width=0.45\textwidth]{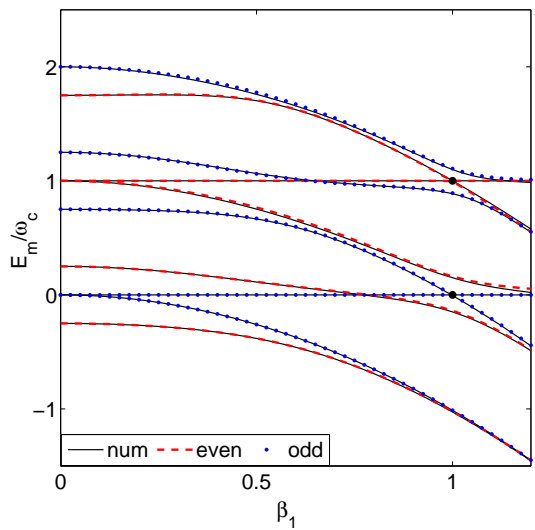}
\caption{(Color Online) The first-order approximation and numerical solution
for the homogeneous coupling system ($g_{1}=g_{2}$) as a function of total
coupling strength $\protect\beta_{1}$ with $\protect\omega _{1}=\protect%
\omega _{2}=0.25\protect\omega _{c}$. The two black points indicate the
level crossing in the same parity subspace.}
\label{Fig4}
\end{figure}

\begin{figure}[t]
\includegraphics[width=0.45\textwidth]{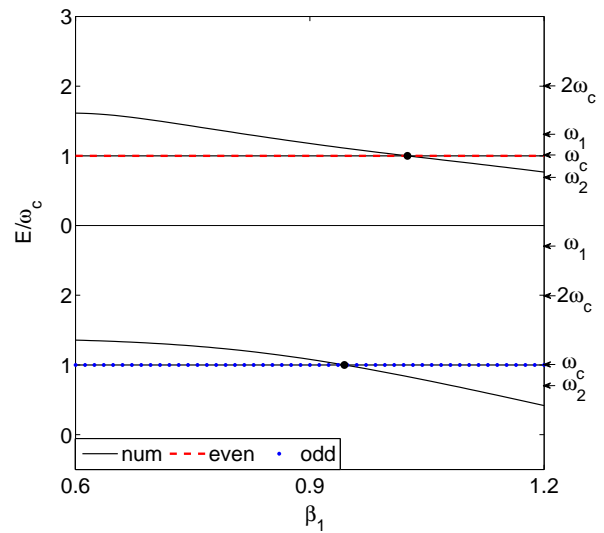}
\caption{(Color Online) Typical level crossing points for homogeneous
coupled non-identical qubits with symmetric detuning $\protect\omega_1+%
\protect\omega_2=2\protect\omega_c$ in the even parity space(upper panel),
and asymmetric detuning $\protect\omega_1-\protect\omega_2=2\protect\omega_c$
in the odd parity space (lower panel). The exact solutions $E=\protect\omega%
_c$ are shown as red-dashed and blue-dotted horizontal lines. The
frequencies of the qubits and the oscillator are labeled on the right side
of the figure.}
\label{Fig5}
\end{figure}

In Fig. \ref{Fig4}, we show the energy spectrum of homogeneous coupling
system as a function of total coupling strength $\beta _{1}$ for $%
g_{1}=g_{2} $ and $\omega _{1}=\omega _{2}=0.25\omega _{c}$. In the
decoupling limit $\beta _{1}=0$, the qubits are set free from the field, and
we find the two odd(even) parity states, i.e. $|01\rangle |m\rangle $ and $%
|10\rangle |m\rangle $, are degenerate for even(odd) $m$, which now means
the photon number of the free bosonic field. The first-order approximation
is valid for the entire region $0<\beta _{1}<1.2$ and it shows that the
energy levels of the same parity would never cross with each other until $%
\beta _{1}\sim 1$. The constant solutions $E_{m}=m\omega _{c}$ are shown as
two horizontal lines $m=0$ for odd parity and $m=1$ for even parity in Fig. %
\ref{Fig4}. Level crossing occurs in the same parity space when other
solutions with different $m$ sweep across them, denoted by two black dots at
$\beta _{1}\simeq 0.9898$ and $1.0004$. The fidelity of both the lower or upper 
states on the two sides of the
crossing points is calculated to be exactly zero, which proves the existence
of the level crossing. In Fig. \ref{Fig4} two parameters $m$ and $\beta _{1}$
are tuned to get these level crossing points in the first-order
approximation, which is consistent with the von Neumann and Wigner
non-crossing theorem.

Besides those shown in Fig. \ref{Fig4}, nontrivial level crossing points may
appear in the fixed parity subspace for non-identical qubits. In the
homogeneous coupling case ($g_{1}=g_{2}$) a level crossing has been found in
the even parity subspace for two inequivalent qubits with symmetric
detunings $\omega _{1}+ \omega _{2}=2\omega _{c}$ \cite{Chilingaryan}
denoted as a black point in the upper panel of Fig. \ref{Fig4}. We here show 
that this is only half of the story. It is actually found that the level crossing in the case of
homogeneous coupling occurs for $\omega _{1}\pm \omega _{2}=2\omega _{c}$.
The reason is that the constant solution $E=\omega _{c}$ holds for either
symmetric or asymmetric detuning conditions. Other levels would inevitably
run across it with increasing coupling strength. In Fig. \ref{Fig5}, we
numerically confirm that the level crossing appears at $\beta _{1}\simeq
1.0251$ for $\omega _{1}=1.3\omega _{c},\omega _{2}=0.7\omega _{c}$ in even
parity subspace \cite{Chilingaryan}, and at $\beta _{2}\simeq 0.9442$ for $%
\omega _{1}=2.7\omega _{c},\omega _{2}=0.7\omega _{c}$ in odd parity
subspace. Level crossings in quantum theory are often related to symmetry.
In the case of inhomogeneous coupling of nonidentical qubits with the
oscillator mode no level crossing in the same parity sapce is found, while $g_{1}=g_{2}$ enlarges the
symmetry so that the level crossing appears on the line $E=\omega _{c}$, and
$\omega _{1}=\omega _{2}$ enlarges the symmetry even further and more
crossing points appear on $E=m\omega _{c}$. This is similar to what happened
in the JC model - the enlarged symmetry by RWA leads to two level crossing
points in the even parity space (Fig. 3 in  \cite{Braak})which are not found in the Rabi model.

The appearance of level crossing and associated symmetry are essentially
related to the integrability of our model, which we shall address in the
following. According to the criterion on quantum integrable proposed by
Braak \cite{Braak}, integrability is equivalent to the existence of $f$
numbers to classify the eigenstates uniquely with $f$ the sum of discrete and
continuous d.o.f. The Rabi model has two d.o.f. and at
the same time can be uniquely labelled by two quantum numbers associated
with the energy level and the parity, respectively. So the Rabi model is
considered to be integrable \cite{Braak, Peng}. We argue that the
homogeneous coupled two-qubit TC model with $\omega_1=\omega_2$ or $%
\omega_1\pm\omega_2=2\omega_c$ is integrable, based on the following three
reasons:

(1) In general a system is not integrable in the whole parameter region, but
it may be integrable under some special conditions. A generalization of Rabi
model with an additional term $\epsilon \sigma_x$ (Eq. (7) of Ref. \cite%
{Braak}) breaking the parity symmetry is constructed to be the first example
of nonintegrable but exactly solvable system. Clearly we recover the
integrable Rabi model for $\epsilon=0$. Another example is spinor
Bose-Einstein condensate of alkali gases \cite{Spinor}. Generally considered
to be nonintegrable, the 1D homogeneous spinor bosons can be exactly solved
with Bethe ansatz (BA) method \cite{Cao} along two integrable lines $c_2=0$
and $c_0=c_2$. In spite of the nonintegrability of our model in the whole
parameter regime, there do exist special situation for the parameters, i.e. $%
\omega_1=\omega_2$ or $\omega_1\pm\omega_2=2\omega_c$, in which case the
homogeneous coupled TC model becomes integrable.

(2) The system is nonintegrable if the total number of d.o.f. exceeds the
quantum numbers used to label the eigenstates uniquely. In the case of
Braak's generalized Rabi model the absence of any level crossing in the
spectral graph is sufficient to rule out its integrability. There are two
d.o.f and one quantum number, energy, is sufficient to number the
eigenstates uniquely. In the case of two-qubit TC model, for the simplified
case of $\omega_2=0$ no level crossing is found in the spectra graph \cite%
{Peng}, while for inhomogeneous coupling accidental level crossing occurs
only for different parities (Figure 1). In both cases the model is
nonintegrable because we have three d.o.f and the two quantum numbers, 
energy and parity, are enough to label the states uniquely.

(3) Level crossing in the same parity subspace is nontrivial because we need
another quantum number other than parity to label the degenerate states,
which leads the system to be integrable. As a typical integrable system,
each eigenstate in the hydrogen atom is assigned three quantum numbers $n,
l, m$ which characterize the quantization of radial, angular and orientation
of the electronic orbit. None of them can be omitted in picking up the state
of this three d.o.f system. This parallels the characterization of each
eigenstate of Rabi model through two quantum numbers, the parity quantum
number $n_0$ and the $n_1$th zero of the transcendental function $G_(x)$,
corresponding to two d.o.f. For JC model with enlarged $U(1)$ symmetry,
level crossing occurs in the same parity space. However, we are lucky that
the operator $C$ can be used for a further decomposition of the subspaces
with fixed parity. The state spaces entails a second possibility to label
the states uniquely through $C$ and a two-valued index $n_0$, with the
parity being a redundant quantum number. Similarly, the level crossing
appeared in homogeneous coupled two-qubit TC model implies an enlarged
hidden symmetry for $\omega_1= \omega_2$ or $\omega_1\pm\omega_2=
2\omega_c$. What we need to do is to find a $C$-like conserved quantity to 
decompose the even and odd subspaces further. Consequently we would have 
three quantum numbers (parity, two-valued index $n_0$ and $C$-like number) 
to uniquely label the state with three d.o.f. Though not an exact result, in the 
zeroth and first order approximation we have shown a possible scheme of labeling 
the states with parity, $\pm $ and $m$. In summary the homogeneous coupled 
model is integrable for two identical qubits or with (a)symmetric detuning, though further
exploration of the conserved quantity and hidden symmetry is needed.

\section{Population inversion dynamics}

To better learn the quantum behavior in the prototypical problem of cavity
electrodynamics with more that one qubit involved, we study the dynamical
properties of a two-qubit system strongly coupled to a high- frequency
quantum oscillator. The eigenvectors and eigenvalues of the system derived
in Sec. II can be taken as a complete set, upon which the time evolution of
wave function can be expanded. We discuss here the probability of finding
the two qubits remaining in the initial state \cite{Irish1}, which is
essentially the fidelity between the wave function at subsequent time $t$
and the initial state.

The simplest dynamical behavior is considered when we put the qubits
initially in any one of the four product states and the initial state of the
oscillator is prepared in the displaced Fock basis corresponding to them. In
the zeroth-order approximation, these initial states are inversely linear
combinations of the eigenvectors of the Hamiltonian (\ref{function}) and
expressed respectively as following
\begin{eqnarray*}
\left\vert 11\right\rangle \left\vert m\right\rangle _{A_{1}} &=&\frac{1}{%
\sqrt{2}}\sum\limits_{\kappa ,\gamma =\pm }d_{1m}^{\kappa \gamma }\left\vert
\psi _{m}^{\kappa \gamma }\right\rangle \\
\left\vert 10\right\rangle \left\vert m\right\rangle _{A_{2}} &=&\frac{1}{%
\sqrt{2}}\sum\limits_{\kappa ,\gamma =\pm }d_{2m}^{\kappa \gamma }\left\vert
\psi _{m}^{\kappa \gamma }\right\rangle \\
\left\vert 01\right\rangle \left\vert m\right\rangle _{A_{3}} &=&\frac{1}{%
\sqrt{2}}\sum\limits_{\kappa ,\gamma =\pm }(-1)^{m}\kappa d_{2m}^{\kappa
\kappa \gamma }\left\vert \psi _{m}^{\kappa \gamma }\right\rangle \\
\left\vert 00\right\rangle \left\vert m\right\rangle _{A_{4}} &=&\frac{1}{%
\sqrt{2}}\sum\limits_{\kappa ,\gamma =\pm }(-1)^{m}\kappa d_{1m}^{\kappa
\kappa \gamma }\left\vert \psi _{m}^{\kappa \gamma }\right\rangle ,
\end{eqnarray*}%
which in the special case of two completely identical qubits reduce to
\begin{eqnarray*}
\left\vert 11\right\rangle \left\vert m\right\rangle _{A_{1}} &=&\frac{1}{2}%
\left( \sqrt{2}\left\vert \psi _{2}\right\rangle +\left\vert \psi
_{3}\right\rangle +\left\vert \psi _{4}\right\rangle \right) \\
\left\vert 10\right\rangle \left\vert m\right\rangle _{A_{2}} &=&\frac{1}{2}%
\left( \sqrt{2}\left\vert \psi _{1}\right\rangle -\left\vert \psi
_{3}\right\rangle -\left\vert \psi _{4}\right\rangle \right) \\
\left\vert 01\right\rangle \left\vert m\right\rangle _{A_{3}} &=&-\frac{1}{2}%
\left( \sqrt{2}\left\vert \psi _{1}\right\rangle -\left\vert \psi
_{3}\right\rangle -\left\vert \psi _{4}\right\rangle \right) \\
\left\vert 00\right\rangle \left\vert m\right\rangle _{A_{4}} &=&-\frac{1}{2}%
\left( \sqrt{2}\left\vert \psi _{2}\right\rangle +\left\vert \psi
_{3}\right\rangle +\left\vert \psi _{4}\right\rangle \right) .
\end{eqnarray*}%
As an example, we study the system dynamics with only one qubit, say qubit
2, being excited to the upper level, i.e. $\Psi \left( 0\right) =\left\vert
10\right\rangle \left\vert m\right\rangle _{A_{2}}$. The probability of
finding the two qubits in any possible product states is easily obtained and
we are interested in the fidelity to the initial state
\begin{equation}
P_{10}\left(m, t\right) =\left\vert _{A_{2}}\left\langle m\right\vert
\left\langle 10|\Psi \left( t\right) \right\rangle \right\vert ^{2}
\label{FP}
\end{equation}%
with
\begin{equation*}
\Psi \left( t\right) =\frac{1}{\sqrt{2}}\sum\limits_{\kappa ,\gamma =\pm
}d_{2m}^{\kappa \gamma }\left\vert \psi _{m}^{\kappa \gamma }\right\rangle
e^{-iE_{m}^{\kappa \gamma }t}.
\end{equation*}%
It is easy to show that $\xi _{m}^{\kappa +}\xi _{m}^{\kappa -}=-1,\left(
d_{1m}^{\kappa +}\right) ^{2}=\left( d_{2m}^{\kappa -}\right) ^{2}$. By
means of these, we find the probability of the two qubits staying in their
initial states consists of four oscillating terms, the frequencies of which
are all possible combinations of $\theta _{m}^{\pm }$, i.e.
\begin{eqnarray}
&&P_{10}\left(m, t\right) =\frac{1}{2}\left\{ 1+\sum\limits_{\kappa =\pm
}\left( c_{1m}^{\kappa }\right) ^{2}\left( \cos \left( 2\theta _{m}^{\kappa
}t\right) -1\right) \right.  \notag \\
&&+\left. \left( 1-c_{2m}\right) \cos \left( \theta _{m}^{+}-\theta
_{m}^{-}\right) t+c_{2m}\cos \left( \theta _{m}^{+}+\theta _{m}^{-}\right)
t\right\}  \notag \\
&&  \label{population1}
\end{eqnarray}%
with the two coefficients defined as%
\begin{eqnarray*}
c_{1m}^{\kappa } &=&\frac{\xi _{m}^{\kappa +}}{1+\left( \xi _{m}^{\kappa
+}\right) ^{2}} , \\
c_{2m} &=&\frac{\left( \xi _{m}^{++}\right) ^{2}+\left( \xi _{m}^{-+}\right)
^{2}}{\left( 1+\left( \xi _{m}^{++}\right) ^{2}\right) \left( 1+\left( \xi
_{m}^{-+}\right) ^{2}\right) }.
\end{eqnarray*}
For homogeneous coupling $\beta _{2}=0$ and $\omega _{1}=\omega _{2}$, eq. (%
\ref{population1}) is reduced to%
\begin{eqnarray*}
&&P_{10}\left(m, t\right) =\frac{1}{8}\left\{ 2+\cos \left( 2\Omega
_{m}^{+}t\right) +\cos \left( 2\Omega _{m}^{-}t\right) \right. \\
&&\left. +2\left( \cos \left( \Omega _{m}^{+}+\Omega _{m}^{-}\right) t+\cos
\left( \Omega _{m}^{+}-\Omega _{m}^{-}\right) t\right) \right\},
\end{eqnarray*}
which consists essentially two oscillating terms because $\Omega_m^+=0$ for
even $m$ and $\Omega_m^-=0$ for odd $m$.

\begin{figure*}[t]
\includegraphics[width=0.9\textwidth]{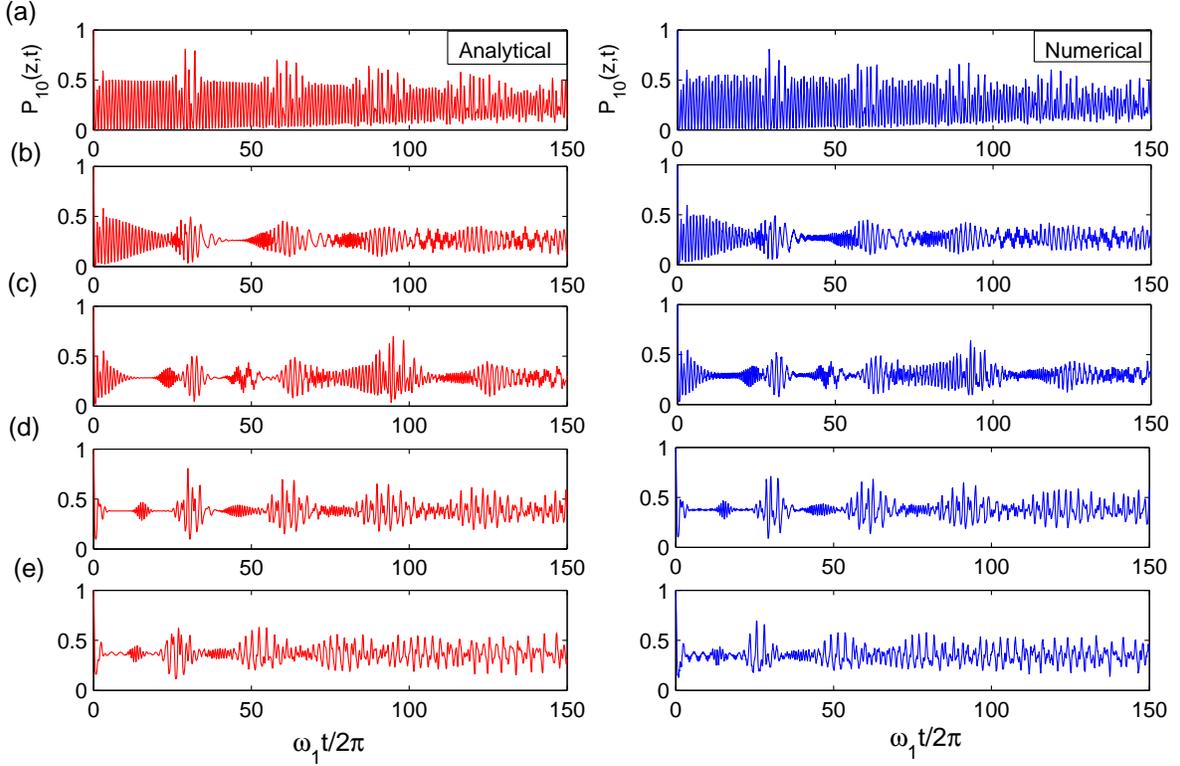}
\caption{(Color Online)Probability $P_{10}(z,t)$ of finding two qubits in
state $\left\vert 10\right\rangle $ as a function of $\protect\omega _{1}t/2%
\protect\pi$, by the zeroth-order approximation analytical method and the
numerical result for different coupling strengths. The zeroth-order
approximation presents a pretty good description of the time-dependent
evolution of the probability. We consider two identical qubits with $\protect%
\omega _{1}=\protect\omega _{2}=0.15\protect\omega _{c}$ and $g_1$ is fixed
to $0.1\protect\omega_c$. The coupling strength to the qubit 2 is $g_2=0.01%
\protect\omega _{c} (a), 0.03\protect\omega _{c} (b), 0.05 \protect\omega %
_{c}(c), 0.1 \protect\omega _{c}(d)$ and $0.12 \protect\omega _{c}(e)$,
respectively. The oscillator is prepared initially in its coherent state
with $z=3$.}
\label{Fig6}
\end{figure*}

Consider now the harmonic oscillator in the state which most closely
approaches the classical limit, that is, we choose the oscillator begins in
the displaced coherent state. The initial state is thus given by
\begin{equation}
\left\vert \Psi \left( 0\right) \right\rangle =\sum\limits_{m=0}^{\infty }%
\frac{e^{-|z|^{2}/2}z^{m}}{\sqrt{m!}}\left\vert 10\right\rangle \left\vert
m\right\rangle _{A_{2}}.  \label{initial state}
\end{equation}%
The probability of two qubits remaining in their initial state $\left\vert
10\right\rangle $ is calculated by tracing over all Fock states of the
oscillator as follows
\begin{equation}
P_{10}\left( z,t\right) =\left\langle 10|Tr_{A}\rho (z,t)|10\right\rangle
=\sum\limits_{m=0}^{\infty }p\left( m\right) P_{10}\left( m,t\right) ,
\label{population}
\end{equation}%
where $\rho (z,t)=|\Psi (t)\rangle \langle \Psi (t)|$ is the density matrix
of the system and the normalized Poisson distribution is defined as
\begin{equation*}
p\left( m\right) \!=\!\frac{e^{-\left\vert z\right\vert ^{2}}\left\vert
z\right\vert ^{2m}}{m!}.
\end{equation*}%
We find that the probabilities of two qubits populating in the four product
states oscillate with the same characteristic frequencies. For the states $%
\left\vert 10\right\rangle $ or $\left\vert 01\right\rangle $ the
oscillation is around an equilibrium position $(1-B)/2$, while for the
states $\left\vert 11\right\rangle $ or $\left\vert 00\right\rangle $ the
oscillation equilibrium is $B/2$ with $B=\sum\limits_{m=0}^{\infty
}\sum\limits_{\kappa =\pm }p\left( m\right) \left( c_{1m}^{\kappa }\right)
^{2}$. For the homogeneous coupling case and $\omega _{1}=\omega _{2}$, we
recover the the analytical result established previously \cite{Agarwal} by
keeping only three terms $l=m,m-1,m-2$ in the summation of $\Omega _{m}^{\pm
}$ (eq. {\ref{Xmn}}) and replacing the Poisson distribution by a Gaussian
one for big enough $|z|$
\begin{equation}
P_{10}\left( z,t\right) =\frac{3}{8}+\frac{1}{2}S\left( t,\omega _{1}\right)
+\frac{1}{8}S\left( t,2\omega _{1}\right) ,  \label{population3}
\end{equation}%
where
\begin{equation*}
S\left( t,\omega _{1}\right) =Re\left[ \sum\limits_{k=0}^{\infty }\bar{S}%
_{k}\left( t,\omega _{1}\right) \right]
\end{equation*}%
and%
\begin{equation}
\bar{S}_{k}\left( t,\omega _{1}\right) =\frac{\exp \left( \Phi _{Re}+i\Phi
_{Im}\right) }{\left( 1+\left( \pi kf\right) ^{2}\right) ^{1/4}}  \label{sk}
\end{equation}%
with%
\begin{eqnarray*}
\Phi _{Re} &=&\frac{-\left( \mu -\mu _{k}\right) ^{2}f\beta _{1}^{2}}{%
2\left( 1+\left( \pi kf\right) ^{2}\right) }, \\
\Phi _{Im} &=&\frac{\tan ^{-1}\left( \pi kf\right) }{2}+\mu -\left\vert
z\right\vert ^{2}\left( \mu \beta _{1}^{2}-2\pi k\right) .
\end{eqnarray*}%
Here we have defined $f=\left\vert z\right\vert ^{2}\beta _{1}^{2},\mu
=\omega _{1}te^{-\beta _{1}^{2}/2},\mu _{k}=2\pi k\left( 1+f/2\right) /\beta
_{1}^{2}$. It is obviously that the revival in $S\left( t,\omega _{1}\right)
$ occurs around each time $\mu =\mu _{k}$, the envelope and the fast
oscillatory of which are determined by the exponential and cosine terms
respectively.

We restrict our discussion to the regime of large detuning $\omega _{c}\gg
\omega _{j}$ and ultrastrong coupling strength $g_{j}\sim 0.1\omega _{c}$.
Fig. \ref{Fig6} shows the results of the time evolution of the probability (%
\ref{population}) by means of the zeroth-order approximation analytical
method compared with the numerically exact solution, where we have made a
cutoff for $m$ to a maximum value $30$ because $p\left( m\right) \approx 0$
for $z=3$ and $m\geq 30$. Our approximated results prove to be unexpectedly
powerful, giving accurate dynamics perfectly in the present experimentally
accessible coupling regime. Here we assume that one of two
qubits reaches ultrastrong coupling regime $g_{1}/\omega _{c}=0.1$, while
the coupling strength to the other qubit $\left( g_{2}/\omega _{c}\right) $
changes from small to large. In Fig. \ref{Fig6}(a) $g_{2}$ is much smaller than
$g_{1}$ and there is no collapse-revival phenomena in the evolution of the
probability. As $g_{2}$ increases, the collapses and revivals emerge
gradually and the first collapse becomes faster and faster and the revival
signal is more and more distinct in Fig. \ref{Fig6}(b-c). In Fig. \ref{Fig6}%
(d) the two coupling strengths are equal, resulting in the most regular
shape in the oscillation of probability and the peaks become periodic in
time. Finally in Fig. \ref{Fig6}(e) when $g_{2}$ is larger than $g_{1}$,
collapses and revivals continue for a while and the oscillation becomes
apparently irregular. Because $\left( c_{1m}^{\kappa }\right) ^{2}<c_{2m}$,
the revivals with smaller amplitude are mainly determined by the second term
of Eq. (\ref{population1}) containing $c_{1m}^{\kappa }$, instead, the
revivals with larger amplitude depend on the third and fourth terms
containing $c_{2m}$. The above analysis and discussions suggest that the
collapse-revival phenomena are sensitive to the coupling strength in the
evolution of the probability, as well as it is periodic only for $%
g_{1}=g_{2} $ and $\omega _{1}=\omega _{2}$. We also find that the
probability of two qubits populating in the four product states exhibit
similar envelope of the revival signal.

\begin{figure}[t]
\includegraphics[width=0.45\textwidth]{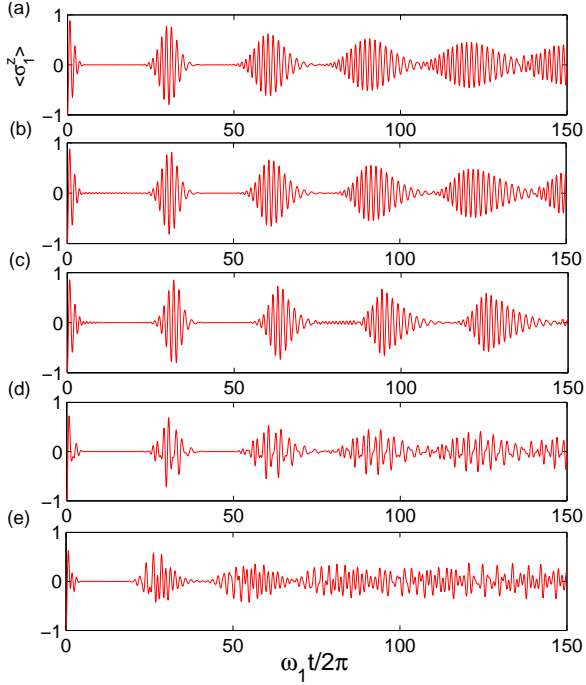}
\caption{(Color Online)The time-dependent inversion of qubit 1 as a function
of $\protect\omega _{1}t/2\protect\pi $ for different coupling strength
between qubit 2 and a bose field, given by our zeroth-order approximation
method. The corresponding parameters are the same as in Figure \ref{Fig6}.}
\label{Fig7}
\end{figure}

It is sometimes more convenient to measure the population inversion in one
of the qubit, that is, we observe the time evolution of the expectation
value of the Pauli matrix operator of qubit 1 defined as $\sigma
_{1}^{z}=\left( \left\vert 1\right\rangle \left\langle 1\right\vert
-\left\vert 0\right\rangle \left\langle 0\right\vert \right) _{1}$ and
related to the probabilities through $\left\langle \sigma
_{1}^{z}\right\rangle =P_{11}(z,t)+P_{01}(z,t)-P_{10}(z,t)-P_{00}(z,t)$. In
doing this, we again fix the value of $g_{1}=0.1\omega _{c}$ and study how
the existence of qubit 2 will change the dynamics of the qubit 1. With the
initial state (\ref{initial state}) the expectation value of $\sigma
_{1}^{z} $ is calculated as%
\begin{eqnarray}
\left\langle \sigma _{1}^{z}\right\rangle &=&\sum\limits_{m=0}^{\infty
}p\left( m\right) \left\{ \left( D_{m}-1\right) \cos \left( \theta
_{m}^{+}-\theta _{m}^{-}\right) t\right.  \notag \\
&&\left. -D_{m}\cos \left( \theta _{m}^{+}+\theta _{m}^{-}\right) t\right\} ,
\label{inversion}
\end{eqnarray}%
with $D_{m}=2c_{1m}^{+}c_{1m}^{-}+c_{2m}$. In the case of $\beta _{2}=0$
this reduces to
\begin{equation}
\left\langle \sigma _{1}^{z}\right\rangle =-S\left( t,\omega _{1}\right) .
\end{equation}%
In Fig. \ref{Fig7} we show the time dependent inversion of qubit
1 for different coupling strength between qubit 2 and the bose field. Due to
the excellent agreement with the numerically exact solution, we only show
the analytical result in Fig. \ref{Fig7} and the parameters are the same as
in Fig. \ref{Fig6}. In the absence of qubit 2, the dynamics of a single
qubit already exhibits the collapse-revival phenomena in the strong coupling
regime $g_{1}\sim 0.1$. The population inversion given in (\ref{inversion})
shows that the revival signal is robust for weak coupling to the second
qubit - we even can not tell the difference for the single qubit dynamics
and that for a coupling strength $g_{2}=0.01$ and the revival signals for
weak coupling cases are only found to be delayed a little in Fig. \ref{Fig7}%
(a-c). With the increasing of $g_{2}$ to the same amplitude as $g_{1}$ the
revival signal is destroyed, indicating that the qubit 2 influences the
qubit 1 by interacting with the optical field.
This behavior can be understood qualitatively as following. For each $m$,
besides a common factor $p\left( m\right) $ Eq. (\ref{inversion}) consists
now of two cosine terms, whose amplitudes are determined by $D_{m}$. For $%
g_{2}<g_{1}$, we can show numerically that $D_{m}$ is always smaller than $%
0.5$ and can be neglected for larger $m$. The dynamics depends thus mainly
on the difference, instead of the summation, of $\theta _{m}^{\pm }$ as in
the first term in (\ref{inversion}). This gives the periodicity of revivals
in Fig. \ref{Fig7}, which would persist even for the homogeneous coupling
case when the two terms in (\ref{inversion}) are comparable. In Fig. \ref%
{Fig7}(e) the interference of the two revival signal terms with almost equal
amplitude leads to the irregular oscillation of population inversion.

\section{Entanglement behaviors}

Quantum entanglement can be used in studies of fundamental quantum phenomena
and the on-chip entanglement of solid-state qubits provides a key building
block for the solid-state realization of quantum optical networks. It has
attracted much attention in connection with Bell's inequality \cite{Bell,
Wootters, Coffman}. However, realization of long-distance entanglement based
on solid-state systems coupled to an optical field is an outstanding
challenge. In the homogeneous coupling case with equal strengths to two
identical qubits the entanglement properties have recently been studied \cite%
{Agarwal}. In this section we aim to describe the entanglement properties by
considering different coupling strengths to the two qubits. Thus it would be
very interesting to study more general quantum correlations between two
qubits. We suppose an initial entanglement of the two qubits in the form of
a familiar Bell state and the oscillator in a coherent state, which is
expressed as
\begin{equation}
\left\vert \Psi \left( 0\right) \right\rangle =\frac{1}{\sqrt{2}}\left(
\left\vert 11\right\rangle +\left\vert 00\right\rangle \right) \left\vert
z\right\rangle .  \label{intial state1}
\end{equation}%
As a good approximations in the case of small $\beta _{1}$ we may expand the
state $\left\vert m\right\rangle$ in terms of the displaced Fock space and
the most important contribution in the summation over $m$ are the terms with
the same $m$, which is equivalent to take $\left\vert m\right\rangle \approx
\left\vert m\right\rangle _{A_{i}}$ \cite{Agarwal}. Thus we can obtain
\begin{equation}
\left\vert \Psi \left( 0\right) \right\rangle \!=\!\frac{1}{\sqrt{2}}%
\!\sum\limits_{m=0}^{\infty }\!\frac{e^{-\left\vert z\right\vert ^{2}/2}z^{m}%
}{\sqrt{m!}}\left( \left\vert 11\right\rangle \!\left\vert m\right\rangle
\!_{A_{1}}\!+\!\left\vert 00\right\rangle \!\left\vert m\right\rangle
\!_{A_{4}}\right) .  \label{initial state3}
\end{equation}%
The initially entangled state of two qubits evolves into $\left\vert \Psi
\left( t\right) \right\rangle $ which is given by%
\begin{equation}
\left\vert \Psi \left( t\right) \right\rangle =\sum\limits_{m=0}^{\infty
}\sum\limits_{\kappa =\pm }\left( e^{-iE_{m}^{\kappa \gamma }t}\left\vert
\psi _{m}^{\kappa \gamma }\right\rangle \left\langle \psi _{m}^{\kappa
\gamma }|\Psi \left( 0\right) \right\rangle \right) .  \label{time state3}
\end{equation}

\begin{figure}[t]
\includegraphics[width=0.45\textwidth]{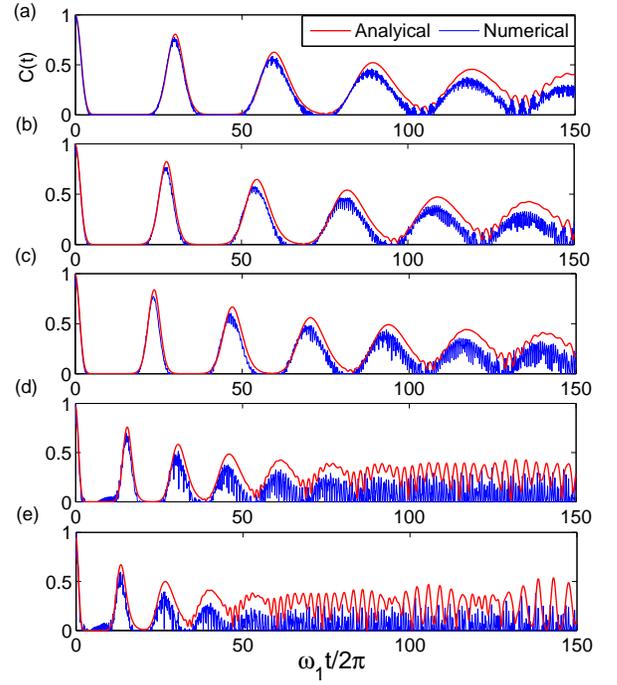}
\caption{(Color Online)Plots of the concurrence evolutions as a function of $%
\protect\omega _{1}t/2\protect\pi $ for different coupling strength between
qubit 2 and a bose field, given by our zero-order approximation approach and
numerical method. The corresponding parameters are the same as in Figure \ref{Fig6}.}
\label{Fig8}
\end{figure}

To quantify the entanglement of a two-qubit system, we need to calculate the
reduced density operator by tracing out the quantum field. The result is
given by
\begin{eqnarray}
\hat{\rho}_{Q}\left( t\right) &=&\sum\limits_{m}\left\langle m|\Psi \left(
t\right) \right\rangle \left\langle \Psi \left( t\right) |m\right\rangle
\notag \\
&=&\sum\limits_{m=0}^{\infty }\sum\limits_{\kappa =\pm }\frac{p\left(
m\right) \left( 1+\kappa \left( -1\right) ^{m}\right) }{4}\left(
q_{1m}^{\kappa +}\left\vert ee\right\rangle \left\langle ee\right\vert
\right.  \notag \\
&+&q_{2m}^{\kappa +}\left\vert ee\right\rangle \left\langle gg\right\vert
+q_{2m}^{\kappa -}\left\vert gg\right\rangle \left\langle ee\right\vert
+q_{1m}^{\kappa -}\left\vert gg\right\rangle \left\langle gg\right\vert
\notag \\
&&  \label{density}
\end{eqnarray}%
where the calculation is done in the eigenbasis $\left\vert e\right\rangle$
and $\left\vert g\right\rangle$ of $\sigma^x$ with eigenvalues $\pm1/2$
respectively. The coefficients $q$'s are defined as
\begin{eqnarray*}
q_{1m}^{\kappa \pm }&=&1\mp \frac{4\xi _{m}^{\kappa +}\left( \left( \xi
_{m}^{\kappa +}\right) ^{2}-1\right) \sin ^{2}\left( \theta _{m}^{\kappa
}t\right) }{\left( \left( \xi _{m}^{\kappa +}\right) ^{2}+1\right) ^{2}}, \\
q_{2m}^{\kappa \pm }&=&1-\frac{8\left( \xi _{m}^{\kappa +}\right) ^{2}\sin
^{2}\left( \theta _{m}^{\kappa }t\right) }{\left( \left( \xi _{m}^{\kappa
+}\right) ^{2}+1\right) ^{2}}\pm i\frac{2\xi _{m}^{\kappa +}\sin \left(
2\theta _{m}^{\kappa }t\right) }{\left( \xi _{m}^{\kappa +}\right) ^{2}+1},
\end{eqnarray*}%
which are all unity at $t=0$. This means that the qubits are initially
prepared in a pure state, but as time evolves, the reduced state of the
qubits becomes mixed. Obviously the reduced density matrix in the eigenspace
of the spin product operators $\sigma_{1}^{x}\otimes \sigma _{2}^{x}$ with
the standard two-qubit basis $\left\vert ee\right\rangle ,\left\vert
eg\right\rangle ,\left\vert ge\right\rangle ,\left\vert gg\right\rangle $
belongs to a special class of density matrices (X-matrices) with only
diagonal and anti-diagonal elements. It is thus more convenient to quantify
the entanglement using concurrence, which in our case takes a very simple
form \cite{Agarwal, Wootters}%
\begin{equation}
C\left( t\right) =\left\vert \sum\limits_{m=0}^{\infty }\sum\limits_{\kappa
=\pm }\frac{p\left( m\right) \left( 1+\kappa \left( -1\right) ^{m}\right) }{2%
}q_{2m}^{\kappa +}\right\vert .  \label{C}
\end{equation}%
It is also worthwhile to mention that for homogeneous coupling $\beta _{2}=0$
we immediately have $q_{1m}^{\kappa \pm }=1$ and $q_{2m}^{\kappa \pm
}=e^{\pm i2\Omega^{\kappa}_m t}$. Then the concurrence (\ref{C}) reduces to
the homogeneous result as in \cite{Agarwal}%
\begin{equation*}
\!C\left( t\right) \!\approx \!\sum_{k=0}^{\infty }\!\left\vert \bar{S}%
_{k}\left( t,2\omega _{1}\right) \right\vert =\!\sum_{k=0}^{\infty }\!\frac{%
\exp \left( \frac{\!-\!\left( 2\mu \!-\!\mu _{k}\right) ^{2}f\beta _{1}^{2}}{%
2\left( 1\!+\!\left( \pi kf\right) ^{2}\right) }\right) }{\left(
1\!+\!\left( \pi kf\right) ^{2}\right) ^{1/4}}.
\end{equation*}

In Fig. \ref{Fig8} we plot the time evolution of the concurrence of
two qubits, coupled to the bose field with different coupling strengthes. It
is interesting to examine how the entanglement changes when one of two
qubits reaches the ultrastrong coupling regime while the other coupling
parameter varies. Obviously, we observe a variety of qualitative features
such as entanglement birth, death, as well as rebirth, in which revivals
appear periodically. Moreover, the periodicity of revivals disappears after
a period of time and the duration of death time becomes shorter and shorter
over time. We realize that the period gets shorter and shorter and the
periodicity of revivals vanishes faster and faster with the increasing of $%
g_{2}$. The zero-order approximation reproduces quite accurate evolution in
short time and fails in describing the long time behavior when the coupling
strengths are sufficiently large.

\section{Conclusion}

In conclusion, we have developed a systematic truncated subspace approach
for solving the TC model beyond RWA by using the displaced Fock basis,
parity operator subspace and truncation in the power series \cite{Irish1, Liu,
Braak}. This provides a straightforward way to access the Hilbert space of 
the inhomogeneous coupling system. In principal we are able to
solve the inhomogeneously coupled $N$-qubits-oscillator model to get an
analytical result to any order by constructing the $2^{N}$ displacement
operators. The complexity of the solutions depends on the determinant of the
secular equation, the primitive building blocks of which involve transition
between Fock states displaced in different directions and distances.
As an example of particular experimental interest, the two-qubit TC model
manifests a lot of new features of the qubit-oscillator system and our main 
findings include:

(1) The analytical energy spectrum 
of the two-qubit inhomogeneous coupling TC model are given in the 
zeroth-order and first-order approximations. The zeroth-order results already agree 
with the numerical solutions very well in the ultra-strong coupling case 
$\beta_{1}\sim 0.2\omega _{c}$, while the first-order approximation improves
the analytical eigenenergies applicable even in the deep coupling regime 
$\beta _{1}\sim \omega _{c}$ after half of the pseudo solutions are ruled out. 

(2) The TC model consisting of two qubits is quasi-exactly solvable, that is,
a finite number of exact eigenvalues and associated eigenfunctions are given 
in the closed form. Specifically, in the homogeneous coupling case, $E=\omega_c$ 
is always a solution corresponding to even(odd) parity for symmetric(asymmetric) 
detuning $\omega_1\pm\omega_2=2\omega_c$. For two completely identical 
qubits homogeneously coupled to the bose field, the singlet state $\left( \left\vert 
10\right\rangle -\left\vert 01\right\rangle \right) |m\rangle /\sqrt{2}$ for any $m$ 
is an exact eigenstate with eigenvalue $E_m = m\omega_c$. The remaining part of 
the spectrum is only numerically accessible through truncation subspace approach.

(3) Several nontrivial level crossing points in the same parity subspace are 
identified by means of the fidelity between states before and after the crossing. 
This implies an enlarged hidden symmetry and we show that the homogeneous 
coupled two-qubit TC model with $\omega _{1}=\omega _{2}$ or $\omega _{1}
\pm \omega _{2}=2\omega _{c}$ is integrable.

(4) The quantum dynamical of the TC model beyond the RWA are investigated in 
the adiabatic approximation, with a special attention paid on the unequal coupling 
strengths for the two qubits. The probability of the two qubits staying in their initial states 
is characteristic of four oscillating frequencies, which is distinct from that of the single 
qubit system and the homogeneous coupling system. The approximated results of 
population inversion are surprisingly accurate in describing the dynamics of the qubit, 
which shows that the collapse-revival phenomena emerge, survive, and
are finally destroyed when the coupling strength increases beyond the deep
coupling regime. This provides a method to control the revival signal of one
qubit by means of the involvement of another one, which imprints its
influences in the system by interacting with the optical field.

(5) The entanglement evolution of the two qubits as a principal measure of
intrinsically quantum coherence is examined with an initial inter-qubit
entanglement in the form of a familiar Bell state and the oscillator in a
coherent state. Analytical results are obtained for the concurrence in the
inhomogeneous coupling case by tracing out the quantum field in the reduced
density matrix. 

Our approximation approach is applicable to systems of
arbitrary two qubits satisfying $\left( \left\vert g_{1}\right\vert
+\left\vert g_{2}\right\vert \right) \leq 0.2\omega _{c}$ and $\omega
_{c}\gg \omega _{j}$. The time evolution of the two qubits reproduces
perfectly the special case with two completely identical qubits
homogeneously coupled to a common oscillator mode, i.e. $g_{1}=g_{2}$ and $%
\omega _{1}=\omega _{2}$ as in Ref. \cite{Agarwal}.
Interestingly, there are still further work to do in the multiple qubits and
oscillator system in the ultrastrong regime, e.g. the GHZ state entanglement
evolution, quantum entanglement between the polarization of a single optical
photon and solid-state qubits, the decoherence behavior analysis in an
external environment, etc.

\begin{acknowledgments}
This work is supported by the NSF of China under Grant Nos. 11234008,
11104171 and 11074153, the National Basic Research Program of China (973
Program) under Grant Nos. 2010CB923103, 2011CB921601. We thank D. Braak, Tao
Liu, Yuxi Liu, Li Wang and Qinghu Chen for helpful discussions.
\end{acknowledgments}

\end{document}